\documentclass[12pt]{iopart}

\usepackage{iopams}
\usepackage{epsfig}
\usepackage{epstopdf}

\newcommand{\celsius}{$^{\circ}\rm{C}$}
\newcommand{\rezA}{\AA$^{-1}$}
\newcommand{\celsiuss}{$^{\circ}\rm{C}$ \,}
\newcommand{\rezAs}{\AA$^{-1}$ \,}

\newcommand{\gcm}{$g \cdot cm^{-3}$}

\newcommand{\Jm}{J/m$^2$}
\newcommand{\Jms}{J/m$^2$\,}
\newcommand{\skut}{CoSb$_3$\,}

\begin{document}

\title[Transport in Au/\skut thin films]{Laser-initiated electron and heat transport in gold-skutterudite \skut bilayers resolved  by pulsed x-ray scattering}

\author{Anton Plech$^1$, Peter Gaal$^2$, Daniel Schmidt$^{2,3}$, Matteo Levantino$^4$, Marcus Daniel$^5$, Svetoslav Stankov$^1$, Gernot Buth$^1$ and Manfred Albrecht$^6$ }

\address{$^1$ Institute for Photon Science and Synchrotron Radiation, Karlsruhe Institute of Technology, Postfach 3640, D-76021 Karlsruhe, Germany, EU}
\address{$^2$ Leibnitz-Institut für Kristallzüchtung, Max-Born-Straße 2, 12489 Berlin, Germany, EU}
\address{$^3$ TXproducts UG, Luruper Hauptstraße 1, 22547 Hamburg, Germany, EU}
\address{$^4$ European Synchrotron Radiation Facility,  71, avenue des Martyrs, CS 40220, F-38043 Grenoble, France, EU}
\address{$^5$ Institute of Physics, Technische Universitat Chemnitz, 09107 Chemnitz, Germany, EU}
\address{$^6$ Institut für Physik, Universitätsstraße 1, 86159 Augsburg, Germany, EU}

\ead{anton.plech@kit.edu}
\vspace{10pt}
\begin{indented}
\item[]February 2024
\end{indented}

\begin{abstract}

Electron and lattice heat transport have been investigated in bilayer thin films of gold and \skut after photo-excitation of the nanometric top gold layer through picosecond x-ray scattering in a pump-probe setup. The unconventional observation of a larger portion of the deposited heat being detected first in the underlying \skut layer supports the picture of ballistic transport of the photo-excited electrons from gold to the underlying layer. The lattice expansion recorded by  x-ray scattering allows accounting for the energy deposition and heat transport.

\end{abstract}

\submitto{\NJP}
\maketitle
\begin{verbatim}
\end{verbatim}
\normalsize
\section{Introduction}

Charge and phonon transport in thin films and nanostructures is of central interest in many functional materials. Among these are certainly microelectronics, but also thermoelectrics, where electron mobility has to be maximized, while phonon heat conduction should be suppressed. Hot-carrier solar cells promise to break the detailed-balance limit \cite{shockley61} in quantum efficiency by selectively harvesting non-thermalized photocarriers \cite{ross82,tagliabue18,su23}. The mobility of electrons is therefore of high importance. 

A standard approach is to model the coupling of heat baths between  electrons, phonons and possibly spin degrees of freedom in coupled differential equations including diffusive spatial transport. This approach is realized in the so-called two-temperature model (TTM) that can be expanded to more degrees of freedom. In the TTM electrons and phonons are considered as two individually thermalized ensembles that exchange energy via electron-phonon coupling $\Delta E_{el.} \propto G(T_e - T_{ph})$ on a time scale of few picoseconds or faster with the coupling parameter G and electron and phonon temperatures $T_e$ and $T_{ph}$, respectively.
The coupling, or electron cooling time $\tau_{e-ph} = \gamma/G \cdot (T_e - T_{ph})$ is proportional to the difference in electron and phonon temperature with $\gamma$ being the electron heat capacity prefactor, determining the slope of heat capacity change with electron temperature. On such short time scales, both systems can also undergo diffusion to transport energy away from a localized heat source. First, upon photo-excitation, the electronic system requires a time duration related to electron-electron coupling for reaching a local thermal equilibrium. Before that electrons can propagate as well. While there is ample documentation of electron and phonon interaction and transport in thin films or nanostructures, ballistic transport is less easy to characterize. Non-thermal or ballistic electrons can move tens of nanometres at speeds of 10$^6$ m/sec through condensed matter (given by the Fermi velocity) while transporting energy faster and further than by thermal electron diffusion. Hohlfeld \etal \cite{hohlfeld00} have studied the ballistic transport of laser-excited electrons in gold and found non-thermal transport over 100 nm. The ballistic carriers were found to be localized in the 6s-p band without contributions from the 5d band of gold. Pudell \etal \cite{pudell18} studied the energy exchange between femtosecond excited gold and adjacent nickel films by using  an ultrafast x-ray diffraction probe. Due to the large difference in electron-phonon coupling between the ultrathin layers (5 nm for gold and 5.6-12.4 nm for nickel, respectively) the electrons in both layers quickly equilibrate to selectively heat nickel independent of laser wavelength (400 nm or 800 nm). The importance of phonon heat transport and thermal interface resistance is pointed out. Ballistic or "superdiffusive" carrier motion is inferred, but can not be distinguished from thermal electron diffusion due to the ultralow thickness of the films.

Liu \etal \cite{liu05} find oscillatory features during femtosecond excitation of 30-45 nm gold films excited from the glass substrate side, which is interpreted as evidence for ballistic electron motion and reflection at the free surface without prior thermalization. The plasmon-polariton-type excitation at the Fermi edge in the d band is characterized by long thermalization times of hundreds of femtoseconds. Ballistic electrons are as well inferred from coherent phonon motion \cite{lejman14} or transport in nanostructures \cite{du13,minutella17,karna23}.

On the other hand, Cahill \etal \cite{choi14,jang20} study the energy exchange between thin metal layers of Au/Pt and Pt/Ru by front side and back side optical probing (time domain thermal reflectivity). They are able to derive values for the electron interface resistance. Furthermore, the TMM describes the thermal dynamics of the Au/Pt system very well and highlights the importance of electron heat transport in systems with dissimilar electron-phonon coupling \cite{wang12,choi14}. 

Skutterudites like \skut represent an interesting material class due to their thermoelectric properties \cite{li19,pang24}. The crystal structure of  \skut shows a void in the cubic unit cell that can host guest atoms, which can modify the phonon thermal transport through localized ratting modes \cite{daniel16}. He et al. have seen that such a reduction of thermal conductivity is manifested in a change of speed of sound of coherent vibrations \cite{he14}.

We investigate the picosecond optical excitation of bilayers of gold as photon absorber and  \skut by time-resolved x-ray scattering. The powder scattering of the crystalline layers serves as a time-resolved thermometer \cite{bracht14,plech19nanomat}. The time resolution of the experiment of 80 ps only captures the thermal expansion, but not coherent phonon motion. We show that the energy partitioning between these films is governed by heat transport through electrons, which is aided by the low electron-phonon coupling factor of gold. The wavelength dependence of energy partition at interband (5d to 6s-p) versus intraband (6s-p) excitation suggests a contribution from non-thermal electrons.

\section{Materials and Methods}

{\bf Thin film growth} The \skut films have been grown on thermally oxidized silicon (100) in a molecular beam epitaxy chamber with Co supplied by an e-beam evaporator and Sb from an effusion cell by co-deposition. The films have been annealed at 500 \celsiuss under UHV conditions. More details can be found in \cite{daniel14}. Gold films were sputter-deposited on top of the grown films.

\begin{figure}
\begin{center}
\epsfig{file=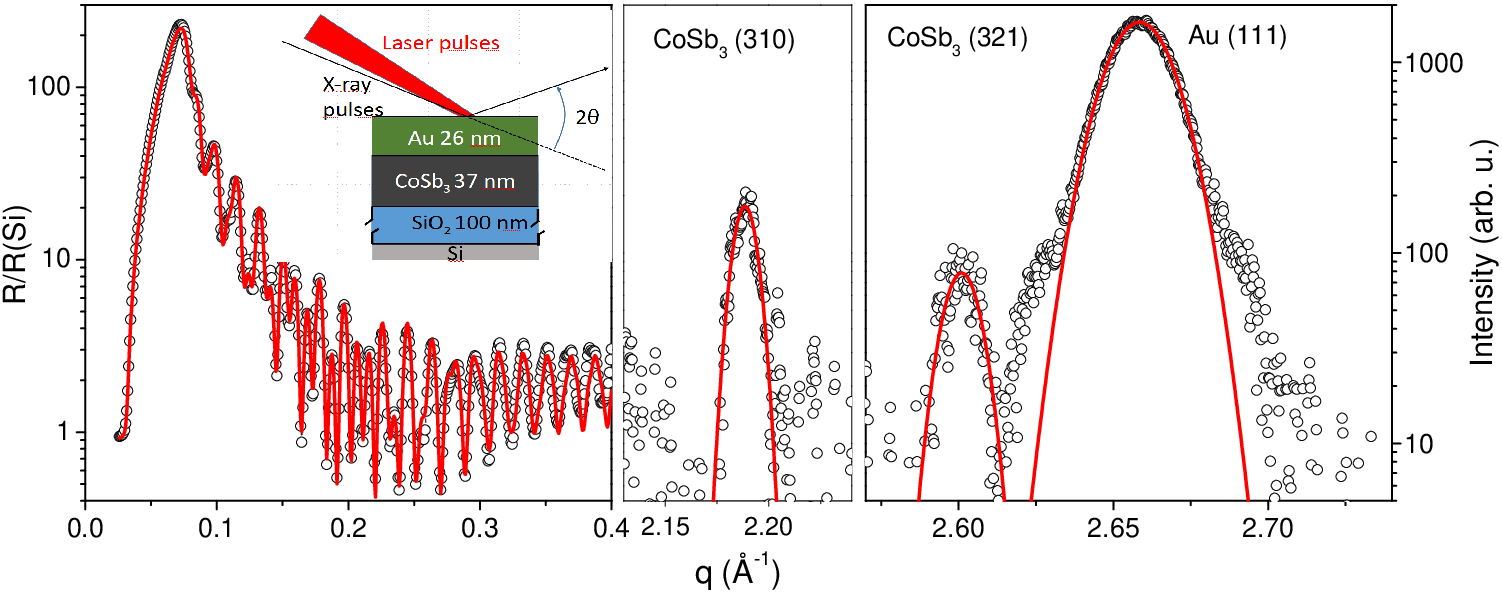,width=15cm, clip=} \caption{From left: X-ray reflectivity of a bilayer of 26 nm gold and 37 nm of \skut on 100nm SiO$_2$ on silicon. The reflectivity is normalized by the reflectivity of the silicon substrate. Powder scattering profile of the \skut (310) peak together with a Gaussian fit and powder scattering profile of the \skut (321) peak at 2.59 \rezAs and the gold (111) peak at 2.665 \rezAs both with a Gaussian fit.}
 \label{fig1}
 \end{center}
 \end{figure}

{\bf Static X-ray characterization} The films were analysed on a six-circle diffractometer at the KIT Light source (Karlsruhe). In brief, a Si(111) monochromatized x-ray beam  with a height perpendicular to the surface of 0.1 mm and a width of 1 mm at 8.9 keV  photon energy was reflected off the surface. The scattered radiation was detected by a line detector (Mythen, Dectris) with 50 $\mu$m pixel size and 1280 pixels. At grazing angles the true reflectivity was obtained after subtraction of the in-plane diffuse scattering from the specularly reflected beam. The reflectivity data was modelled by a multi-layer model using the Parratt algorithm and the software package GenX \cite{genx}. The x-ray reflectivity is shown in fig. \ref{fig1} as normalized by the reflectivity of the silicon substrate with a roughness of 3 $\AA$. The simulation reveals a \skut thickness of 37 nm and a gold thickness of 26 nm. The interfacial roughnesses were in the range of 7-8 $\AA$. Powder scattering was recorded in the same geometry with the incidence angle fixed at the nominal Bragg position of gold and \skut, respectively. In fig. \ref{fig1} two peaks from \skut are discerned, the (310) reflection around 2.2 \rezAs and (321) reflection at 2.59 \rezA. The gold (111) reflection is seen in the same detector exposure as the (321) reflection and shows a strong texture in $<$111$>$ direction perpendicular to the surface \cite{plech19nanomat}. Although the sputtering process produces a poly-crystalline film the crystallographic preferential orientation produces a strong (111) texture of the gold layer. The vertical grain size approaches the film thickness, which produces Laue oscillations, as seen in fig. \ref{fig1}. By heating the wafer steadily by an attached resistive heater  (controller Lakeshore 320) and a controlled temperature ramp the shift in Bragg position can be recorded to deduce the lattice expansion, which is almost linear in the range from 25 to 150 \celsius. The linear expansion coefficient for \skut was found to be 8.3$\cdot$10$^{-6}$, while that of gold was determined to be 18.3$\cdot$10$^{-6}$. These values are larger than the tabulated bulk expansion coefficient \cite{touloukian} because in thin-film geometry the lateral expansion of the film in the surface plane is restricted. Due to the Poisson effect the perpendicular expansion thus compensates with a larger expansion \cite{plech19nanomat}, see Appendix A.

{\bf Time-domain X-ray thermal scattering} Lattice dynamics of the pulsed-laser heated films were recorded at the beamline ID09 at the European Synchrotron Radiation Facility (ESRF, Grenoble). Briefly, a Ti:Sa amplified laser system excites the sample at the fundamental wavelength of 800 nm or the second harmonic at 400 nm of 1 ps length with pulses at 1 kHz. The light is focused by a lens to 1 mm and shone at an 72$^{\circ}$ angle to the surface normal to match the x-ray footprint at an angle of 80 $^{\circ}$ with a size of 0.04 mm. The intensity and polarization are varied by a motorized combination of Glan-Laser prism and wave plate. The x-ray pulses are produced by a single-line undulator \cite{plech02jsr} at 15 keV with a band width of about 1.2 \%  and diluted to the  1 kHz repetition rate by a mechanical chopper \cite{cammarata09}. The scattering from 2000 pulses was integrated on an area detector at 220 mm distance at an incidence angle corresponding to the gold (111) reflection. Images at a given delay (tunable in 5 ps steps, compared to the x-ray pulse length of 80 ps full width at half maximum) were recorded before and after laser excitation. Integration of the images along the radial scattering angle $2\Theta$ allows for extracting the powder profile I(1) of the sample (q $=$2$\pi$/$\lambda \cdot$sin(2$\Theta$)), where the above-mentioned peaks were used to derive the delay-dependent lattice expansion. 

{\bf Theoretical modelling} Pure lattice heat transfer \cite{cahill03} is modelled by a matrix transfer method (called 'transmission line') using a Laplace-domain ansatz \cite{chen99,plech19nanomat}. The basic assumption is that initially a layer on top of a defined layer stack is raised to a new, higher temperature. After that heat is subsequently transferred to the layers beneath through heat conduction perpendicular to the surface (one-dimensional transfer) to be dissipated in the semi-infinite substrate. Importantly, heat transfer between layers is finite due to the thermal impedance between two dissimilar materials. This Kapitza or thermal interface (TIR) resistance is known to be an important factor to the heat dissipation \cite{cahill03,issenmann13,bracht14}. The transmission line approach does not allow to include heat transfer in both directions of the surface normal, but only towards the cold substrate from the topmost heated layer. The free parameters are the thermal conductivities of the layers and substrate and the TIR at each interface. Here we fix to the conductivity of gold, silicon dioxide and the silicon substrate to the tabulated bulk values (see Appendix B), while the \skut  conductivity and the TIR can be varied. 

Coupled thermal electron and lattice heat transport is modelled by a system of coupled differential equations, the so-called two-temperature model (TTM) describing interaction of two heat baths of electrons and phonons and spatial diffusion of heat carried by the electron and phonon systems, respectively. The calculations have been performed using the numerical code package NTMpy \cite{alber21}. This code solves coupled parabolic differential equations in one dimension perpendicular to an excited material surface. The input parameters are film thicknesses, electronic and lattice heat capacities, electronic and lattice heat conduction and the electron-phonon coupling constants for each material. The  energy input (source term) is included by depth-dependent heating of the electron gas according to the refractive indices of the layers with the option allowing for a variable incident angle. It should be noted that no TIR is included in the code. For used material parameters see Appendix B.

\section{Results and discussion}
 
\subsection{Temporal energy distribution between the layers}

The laser pulses are partially absorbed in the top gold layer with nominal penetration depths at normal incidence of 16 nm at 400 nm and 8-12 nm at 800 nm wavelength \cite{bonn00,chen11pnas,fot19}. While the gold layer thickness is considerably larger than the penetration depth of 400 nm and 800 nm light, respectively, at normal incidence, it is useful to inspect the distribution of excitation at the grazing angles used here. For this purpose a simulation of the NTMPy code was modified by calculating the electron temperature across the gold film for a 100 fs laser pulse with all contributions to transport (electron heat conductivity, electron-phonon coupling) switched off. This electron temperature distribution directly after the laser pulse duration should then represent the linear light absorption distribution. The corresponding temperature profiles are shown in Appendix C. The curves are normalized by the maximum derived temperature at the gold surface. The fluence values were 15 \Jms for 400 nm and 150 \Jms  for 800 nm. As, however, possible changes of the refractive index with excitation density are not incorporated in the model, these values should be fluence independent. One obtains an almost exponential decay of the excitation with depth with characteristic values of 39 nm at 400 nm and 18.5 nm at 800 nm. These values are still larger than typical literature values, but agree that most of the laser energy should be deposited first in the gold layer. Caution should be given to the detailed distribution at the gold-skutterudite interface, where the local field distribution may vary. 
Figure \ref{fig8} in Appendix C shows that the electron temperature at the interface reaches 75 \% of the surface temperature for 400 nm and only 26 \% for 800 nm. Therefore, most of the energy in the electron system is initially deposited into the gold layer, amounting for 80 \% of the energy density at 800 nm and 69 \% at 400 nm, assuming an exponential decay of excitation density.  

In any case, the excited electrons would thermalize within about a picosecond \cite{perner97,bonn00} and couple to the phonon bath with electron phonon coupling in gold, which takes about 2 picoseconds at low electron temperatures, but increases with intensity due to a reduced coupling at elevated electron temperatures \cite{hohlfeld00}. In parallel, the hot electrons may diffuse perpendicularly to the surface. 

\begin{figure}
 \begin{center}
\epsfig{file=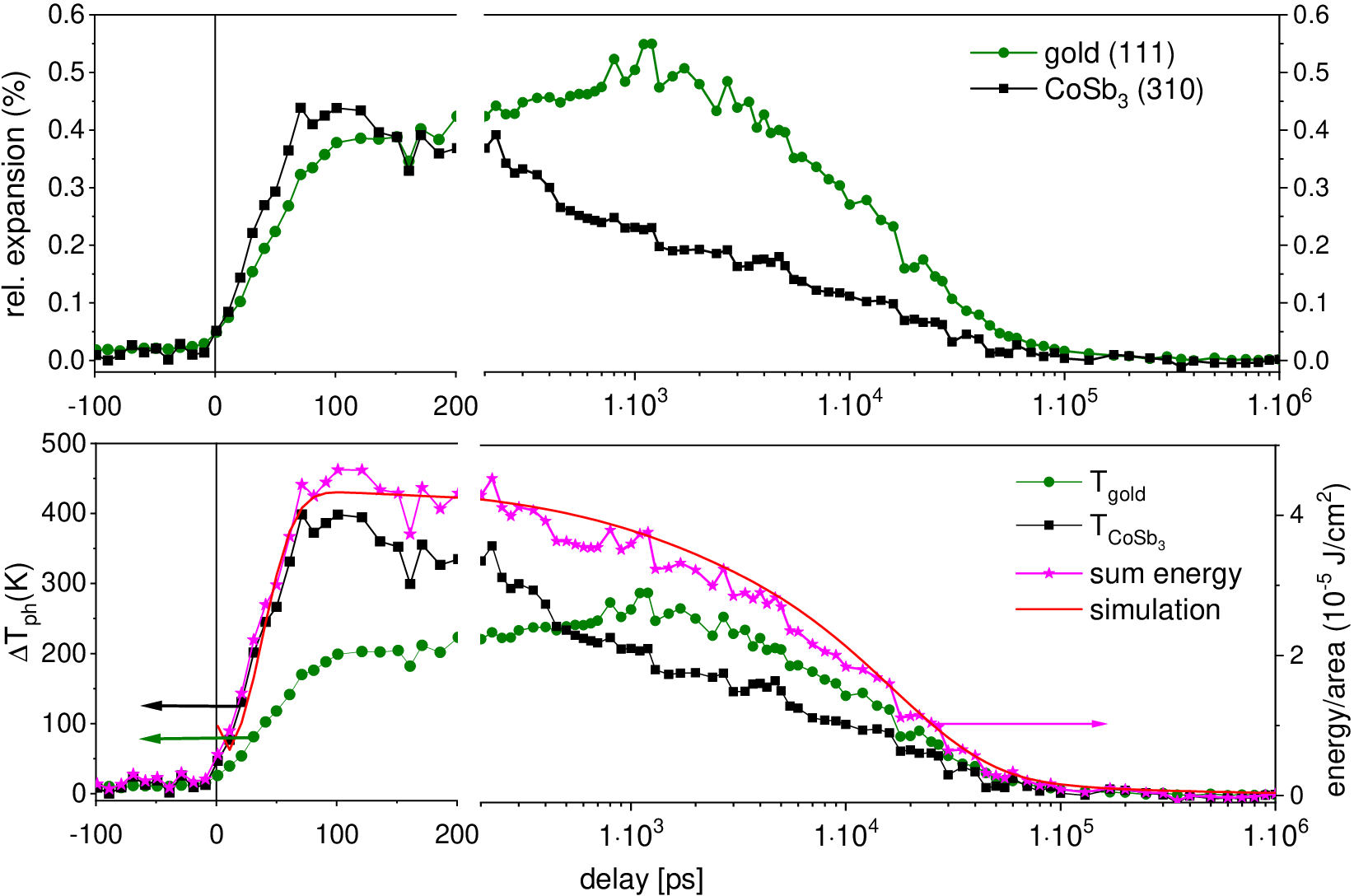,width=15cm, clip=} \caption{Top: Lattice expansion of the gold and \skut layers as function of delay after laser excitation at 800 nm (p polarization). Bottom: Conversion of the lattice expansion into temperature as function of delay of the two layers (left y axis) together with the integral thermal excess energy of both layers after laser excitation. The line represents a simulation of energy change during heat transfer if both layers were heated equally. Measurement errors are represented by the high-frequency noise of the data.}
 \label{fig2}
 \end{center}
 \end{figure}

The lattice expansion of the layers is shown in fig. \ref{fig2} at an incident laser fluence of 28.6 \Jm at 800 nm and p polarization. Both layers of Au and \skut expand at the same time upon arrival of the laser (picosecond delay). The rise time of the expansion signal is determined by the x-ray pulse length of about 80 ps. Within this time the layers reach an expansion of about 0.4 \%. At larger delays one can observe that the expansion of \skut decreases, while that of gold increases further up to a value of 0.5 \% at 1 ns delay, after which it also decreases. At 100 ns both layers have reached their initial lattice parameter. This shows that cooling is completed within 100 ns.

 Given that the laser energy is mostly absorbed in the gold layer it is counter-intuitive that the buried \skut layer even expands further. The lattice expansion can be directly converted into lattice temperature assuming that the layers are internally thermalized within the time resolution of the experiment and a unidirectional expansion normal to the surface of the layers occurs, using the Poisson relationship of a clamped layer with the substrate temperature not changing. This deduction has been shown to be well fulfilled in earlier time-domain x-ray scattering experiments \cite{issenmann13,bracht14}, where was shown that the in-plane expansion of a 200 nm thick gold film on silicon is much smaller as compared to the normal direction of expansion \cite{plech19nanomat}.  In fig. \ref{fig2} b) the expansion is converted into temperature. The temperature increase of the buried \skut layer is even more pronounced than the lattice expansion as compared to that of gold. The reason is the lower expansion coefficient of \skut as compared to gold as derived above. The temperature rise of \skut at 100 ps is twice as high as that of gold. Qualitatively one can deduce an important heat flow from \skut to gold within the first nanosecond, despite the gold layer being selectively excited by the laser. This leads to the conclusion that lattice heat ends in the \skut layer within the first 100 ps by transport of heated electrons. 

The total excess energy of the sum of the two layers upon laser excitation can be calculated by including lattice specific heat and the thickness of the films. This total energy per area is shown in fig. \ref{fig2} bottom marked by stars. It increases within 100 ps to drop later on a 1-10 ns time scale. This behaviour is indeed what would be expected by a impulsively heated layer that cools down by contact with the substrate. A modelling of the heat transfer by the transmission line approach (solid line in fig. \ref{fig2} bottom) indeed reaches a good match to the measured signal assuming that the heat is being input in both layers at the same time with bulk material parameters \cite{plech19nanomat}. The interface resistance found for the \skut - SiO$_2$ is 2.6$\cdot$10$^7$ m$^2$K/W, which is a typical value for the impedance mismatch between dissimilar materials \cite{cahill03,bracht12,issenmann13}. Thus, it is reasonable to assume that heat transfer to the SiO$_2$ layer and the silicon substrate is given by phonon diffusion as would be expected. 

 \begin{figure}
 \begin{center}
\epsfig{file=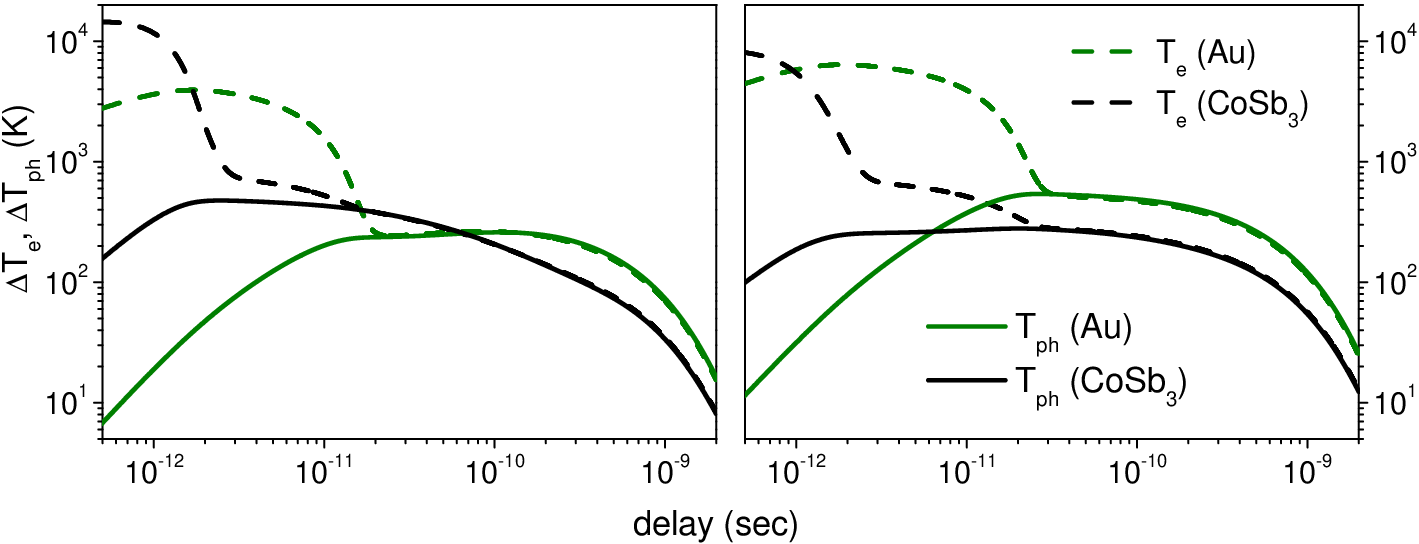,width=15cm, clip=} \caption{Output of the TTM simulation for both 800 nm (left) and 400 nm (right) laser wavelength. The electron temperatures of \skut and gold are indicated by dashed lines, whereas the lattice temperatures are shown as full lines. The different colours green and black represent the mean values of the gold and \skut layers, respectively. The laser excitation event has been deliberately shifted to 1 ps with 1 ps long pulses for clarity. }
 \label{fig3}
 \end{center}
 \end{figure}
 
The energy partition between gold and \skut, on the other hand, is counter-intuitive. Therefore, it can be concluded that energy transport across the gold-\skut interface is done by excited electrons before the lattice is given time to expand. Note that in principle, phonon excitation in gold could show vibrational dynamics beyond the present time resolution with a characteristic period around 14 ps \cite{issenmann13}. This would, however, exclude that energy flows back into the gold layer for as long as 1 ns. It is indeed reasonable to assume that electrons can move away from the gold layer and couple to the \skut lattice to heat the buried layer directly. Assuming that thermalized electrons dissipate via diffusion one can evaluate the TTM to find the electron and lattice temperatures as a function of delay for each layer individually. These quantities have been simulated by the code NTMPy and presented in fig. \ref{fig3} for both excitation at 800 nm and 400 nm.  It can be seen that the initial electron temperature in \skut rises to values of up to several 1000 K, while the electron temperature of gold stays initially lower (dashed lines in fig. \ref{fig3}), but the \skut electron temperature drops faster, which is countered by a fast rise of \skut lattice temperature. On the other hand, both the lattice temperatures reach similar values after the electron gas cools down at 10-20 ps. The reason is the slow electron-phonon coupling in gold, which allows the electrons to diffuse towards the \skut layer. The lattice temperatures of gold and \skut are always considerably lower than the electron temperatures because of the strongly differing heat capacity. The excitation fluence has been set to match the observed expansion in fig. \ref{fig2} at 800 nm. At 400 nm a lower fluence has been used to compensate for the lower reflectivity of the gold layer below the plasma wavelength.

The simulation model does predict that the lattice temperature of the \skut layer rises above that of the gold layer for 800 nm, but not for 400 nm. After a delay of 100 ps for 800 nm and 10 ps for 400 nm both layers have thermally equilibrated. In the experiment, on the other hand, equilibration takes some 500 ps. The faster time scale in the simulation is likely related to the negligence of TIR across the gold-\skut interface.

\subsection{Dependence on wavelength and polarization}

\begin{figure}
 \begin{center}
\epsfig{file=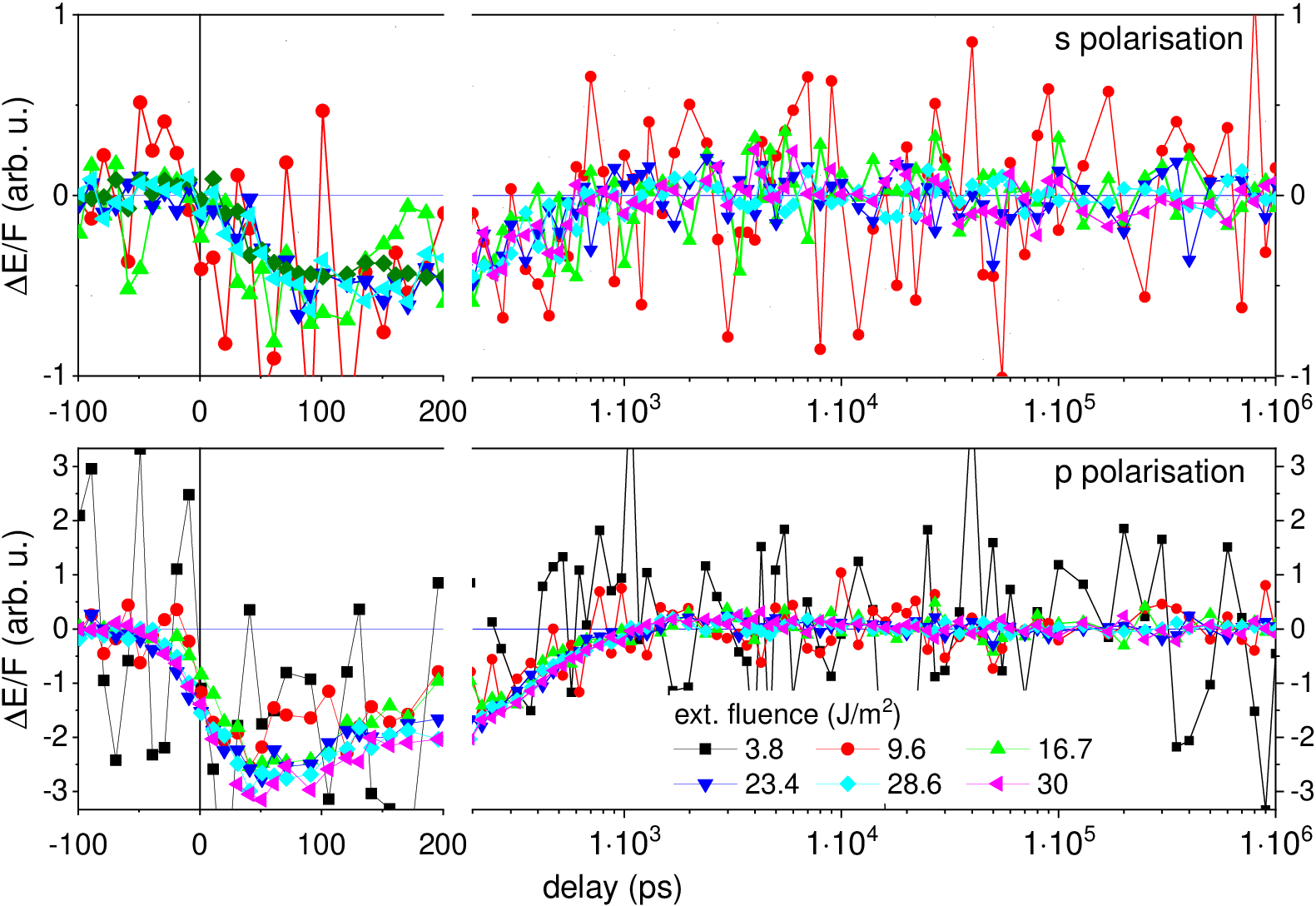,width=15cm, clip=} \caption{Net energy dissipation per area in arbitrary units from the gold layer as function of delay. Negative energy change shows heating of the gold layer, while positive values show cooling. Data is compiled for a range of increasing fluences for p-polarization (top) and s-polarization (bottom) of the laser pulses at 800 nm.}
 \label{fig4}
 \end{center}
 \end{figure}

One might imagine that excitation at 800 nm can still lead to direct light interaction with the buried \skut layer due to possible surface plasmon-polariton excitation of the gold and/or \skut layer analogous to total internal reflection excitation \cite{homola99}. To shed light on this question we changed laser polarization as well as laser fluence at 800 nm to examine the heat flow between the layers. Fig. \ref{fig4} shows the net heat flow between gold and \skut derived by subtracting the excess heat in the gold layer from that of the \skut layer. The signals are in each case divided by the applied fluence for normalization. Heat flow from gold to \skut leads to a positive value of the energy difference, while the inverse flow direction is characterized by negative values. Indeed, heat flow is always towards the gold layer within the first 500 ps independent of the laser fluence and laser polarization. After 500 - 800 ps the flow turns positive, which is explained by the further cooling of gold towards \skut and SiO$_2$ and substrate in accordance to what has been discussed above. It is noted though, that the signal is more noisy for s polarization and reaches lower extreme values than for p polarization. The reason is simply that the s-polarized laser beam at oblique angles displays a higher reflectivity according to the Fresnel equations than that for p polarization. The weak dependence of the heat flow on laser fluence shows that the process is mainly linear, not involving non-linear effects. Note that the data for the lowest fluence is in all cases quite noisy due to the fact that the heating and thus lattice expansion approaching the detection limit of about 2 K \cite{plech19nanomat}. Concluding, we find an initial heat flow towards the gold layer for 800 nm and at high fluence also for 400 nm, but no polarization dependence in any case, which would imply polaritonic modes during excitation.

\begin{figure}
 \begin{center}
\epsfig{file=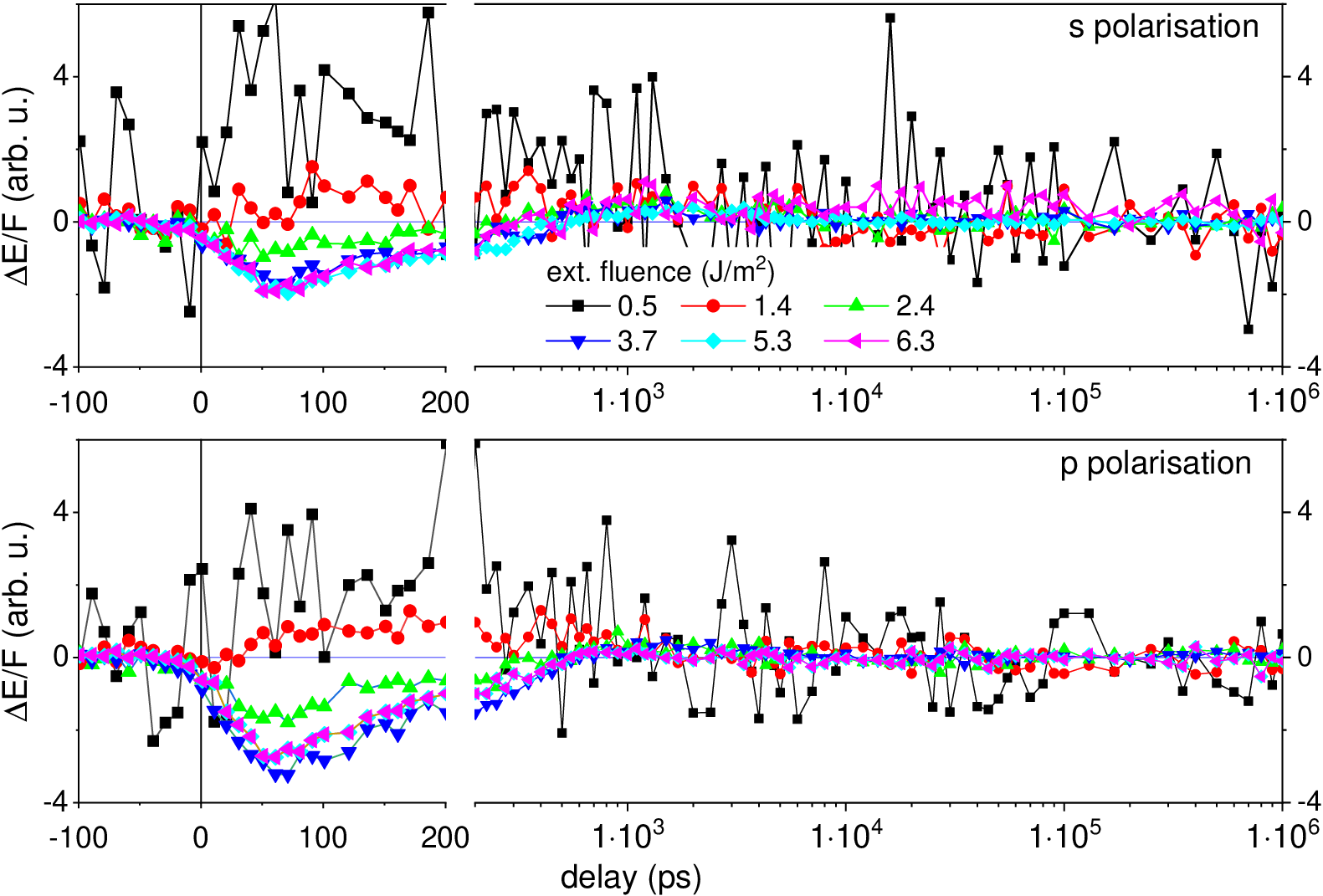,width=15cm, clip=} \caption{Net energy dissipation per area in arbitrary units from the gold layer as function of delay. Negative energy change shows heating of the gold layer, while positive values show cooling. Data is compiled for a range of increasing fluences for p-polarization (top) and s-polarization (bottom) of the laser pulses at 400 nm.}
 \label{fig5}
 \end{center}
 \end{figure}

 This behaviour is different for 400 nm excitation, which is displayed in fig. \ref{fig5}. At 400 nm, for the lowest fluence the heat flow is positive in both settings for polarization, which indicates that the initial lattice heating is localized in the gold layer directly after excitation, where after heat flows towards the \skut layer. The \skut layer does not stay completely cold during the first 100 ps, but accepts some 20 \% of the energy. At higher fluence, on the other hand, the net flow before 500 ps is subsequently reduced before turning negative. This represents the situation that has been seen for 800 nm. At high fluence, the electrons are again able to escape the gold layer to penetrate \skut and be thermalized with the lattice there. 
 
\begin{figure}
\begin{center}
\epsfig{file=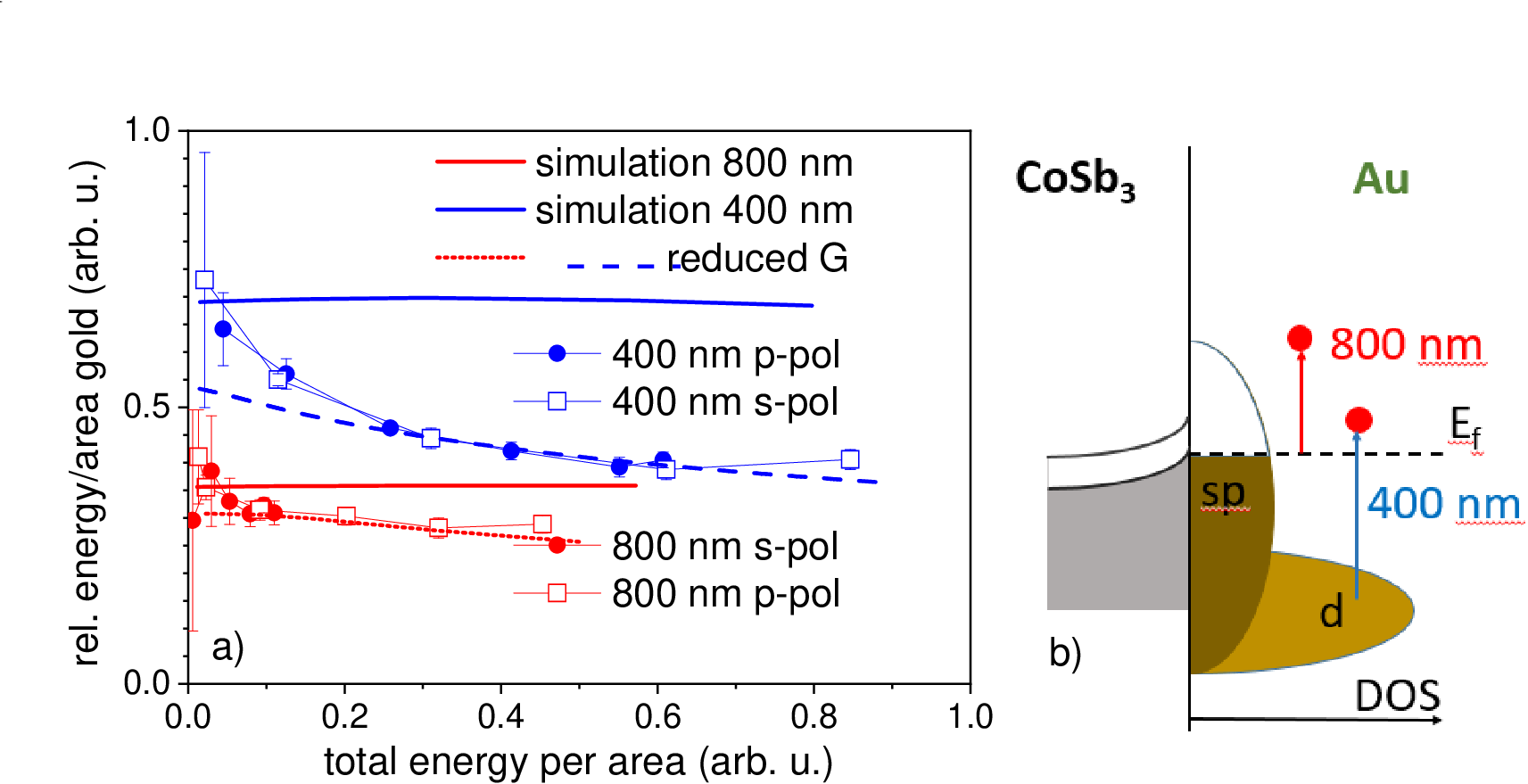,width=15cm, clip=} \caption{Left: Fraction of the energy per area deposited initially (at a delay of 100 ps) in the gold layer relative to the total energy deposition as function of laser fluence for both wavelengths and polarizations. The full lines show the energy fraction of gold as derived from the TTM simulations at a delay of 20 ps for 400 nm and 800 nm at the nominal gold e.-ph. coupling factor of 2$\cdot$10$^{16}$W/m$^3$/K. The dashed lines show the same simulations, but for a reduced coupling factor of 0.35$\cdot$10$^{16}$W/m$^3$/K at 400 nm and 0.6$\cdot$10$^{16}$W/m$^3$/K at 800 nm. Right: Schematic energy band of gold in contact to a semiconducting \skut with the excitation from close to the Fermi edge at 800 nm and from the d band at 400 nm.}
 \label{fig6}
 \end{center}
 \end{figure}

For the purpose of quantifying the eventual  shift in energy deposition as function of fluence, the total energy of the gold layer at 100 ps has been divided by that of the total excess energy in both layers at 100 ps as demonstrated in fig. \ref{fig2} bottom. This ratio represents a measure of the fraction of energy deposited in the gold layer before lattice thermal transport sets in. Fig \ref{fig6} displays this ratio as function of the total excess energy rather than the fluence. Comparing the fluence values is not useful here because of the variable ratio between absorption and reflection with polarization as well as wavelength. 
 
For 800 nm there is not much variation of this ratio, as discussed before. In average only about 30 \% of the laser energy is deposited into the gold layer as lattice heat. Contrarily, at 400 nm at low fluence, a much higher fraction of up to 75 \% is deposited in the gold layer. 

The simulation predicts that 69 \% of the lattice energy is initially (at 20 ps) deposited in the gold layer at 400 nm, but only 30 \% is deposited in gold at 800 nm. This is in agreement with the experiment at low fluence, but does not reproduce the reduction of deposited heat in gold with increasing fluence in the experiment. This reduction is significant for 400 nm with the deposited energy decreasing to 38 \%, comparable to the ratio for 800 nm. In the simulation the coupling constants considered as being independent of wavelength and electron temperature.  

\subsection{Discussion}

The TTM calculations show that a pure diffusional transport of heat by electrons and phonons would not fully account for the preferential heating of the buried \skut layer with the given parameters in particular at 400 nm excitation. Thus, a different explanation needs to be given for this observation. 

A direct explanation for the change in thermal energy deposition in gold as function of fluence is the notion that the cooling time $\tau_{e-ph}$ for the hot electrons would depend on the temperature difference to the phonon bath. Thus, at higher fluence as the electron temperature increases the cooling time would increase accordingly. During this prolongation of a population of heated electrons the effect of transport from gold to the \skut layer would increase, thus reducing the relative phonon heat deposited in the gold film. 

On the other hand, such an effect is not predicted in the simulation, when applying the reported values for the coupling factor G in gold of  2$\cdot$10$^{16}$W/m$^3$/K. It is, on the other hand, possible to reproduce a reduction of the relative heating of gold by reducing G to extremely low values of 0.35 and 0.6$\cdot$10$^{16}$W/m$^3$/K for excitation at 400 nm and 800 nm, respectively. It should be emphasized that such a low value of G has not been reported. 

\begin{figure}
\begin{center}
\epsfig{file=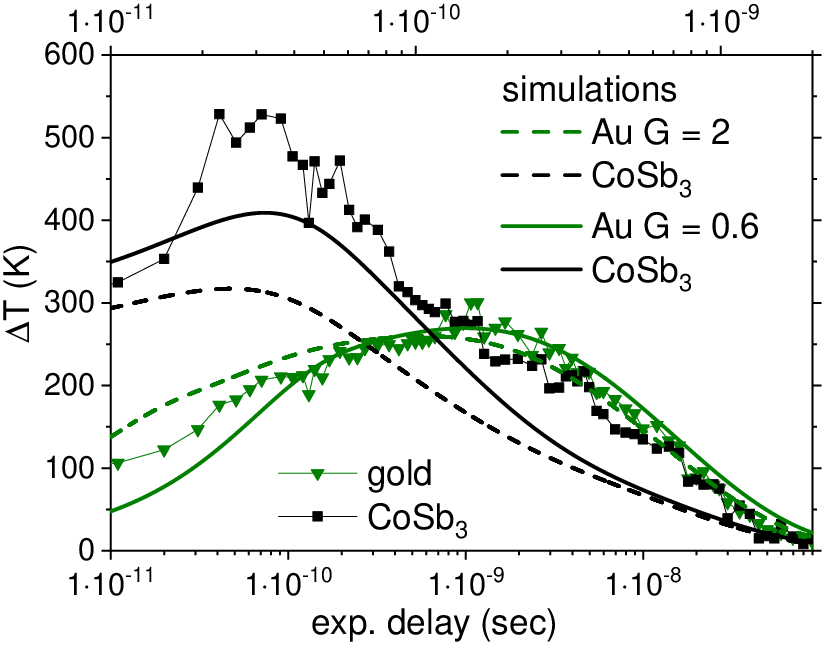,width=12cm, clip=} \caption{Comparison of experimental lattice heating with TTM simulations of the transient lattice temperature change of the gold and \skut layer G$=$ 0.6$\cdot$10$^{16}$W/(m$^3$K) at 800 nm,  but with a scaled delay axis. The simulations results are compared the measured lattice temperatures as displayed in fig. \ref{fig2} with G$=$ 2$\cdot$10$^{16}$W/(m$^3$K).}
 \label{fig7}
 \end{center}
 \end{figure}

A real difference in G for the used wavelength could be present, but is difficult to rationalize. Chiang \etal \cite{chiang23} show that the intrinsic e-ph relaxation time in gold nanoparticles of 40 nm diameter is independent of wavelength, when extrapolating the excited T$_e$ to zero. On the other hand, while acknowledging the applicability of the TTM Minutella \etal \cite{minutella17} find different slopes of the relaxation time above the interband energy and below, which is linked to the coupling factor. Hopkins, on the other hand, shows that interband and intraband excitation leads to different responses as characterized by differing electron heat capacity and electron-phonon coupling \cite{hopkins10}.

Hohlfeld \etal \cite{hohlfeld00} caution that excitation at 400 nm would change the transient reflectivity of a thin gold film less that at 800 nm due to the lower sensitivity to the electron temperature. In our case the 1 ps laser pulses would reduce this effect due to the lower peak power. Also, monitoring the maximum total energy in the layers as function of applied external fluence yields a linear relationship, which excludes non-linear self-interaction of the pump laser.

We conclude that the apparent discrepancy in the magnitude of the electron phonon coupling in the simulation in fig. \ref{fig6} and the well documented couplings constants in literature are rooted in the negligence of ballistic electron transport across the interface. Non-thermal or ballistic electrons would travel at a speed close to the Fermi energy, which allows them to cross the gold layer before being thermalized. If the gold-skutterudite interface would show no Schottky barrier, these electrons would be able to enter the \skut layer to be stopped only at the \skut -SiO$_2$ interface. Low electron conductivity and a high electron-coupling factor in \skut would then transfer the energy efficiently to the lattice. On the other hand, the strong effect with fluence at 400 nm, i. e. the very low apparent G would be rooted in the difference in electron temperature between excitation at 800 nm an at 400 nm \cite{minutella17}. At 800 nm, the electrons are excited directly from or close to the Fermi energy transforming the photon energy into electron kinetic energy. At 400 nm, on the contrary, the excitation from the low-lying d band leads to a lower distance of the electron energy from the Fermi edge, with more energy stored in the excitation of holes. This is sketched in fig. \ref{fig6}. These holes would have a lower mobility, thus not carrying energy into the \skut layer. Given that the relaxation time scales with electron temperature, the cooling at 400 nm would be faster, but would slow down with higher fluence. This effect only plays out in the right regime of G, which needs to be lower than the reported values. This apparent reduction in G is in reality a contribution of ballistic transport. Such difference in non-thermal electron distribution is discussed in \cite{tagliabue18}. In fact, the mean free paths (MFP) of conduction band electrons is considered to be at least 100 nm \cite{hohlfeld00}, with some reports finding even longer MFP of 150 nm \cite{karna23}.

Finally, the experimental temperature curves are compared to the simulations in fig. \ref{fig7} by introducing the apparent reduced G values.  In order to match the experimental resolution a Gaussian convolution matching the x-ray temporal resolution of 80 ps has additionally been applied to the simulations. The electron phonon coupling has been reduced to G$=$ 0.6$\cdot$10$^{16}$W/(m$^3$K) in order to reproduce the significant rise of the \skut layer temperature at early times of 100 ps before the heating of the gold layer as derived from the experiment at 28 \Jm. 

The absolute time scales are shifted in the experimental result with respect to the simulations, which we attribute to not including the interface resistances for phonon heat and possibly electron diffusion \cite{jang20}. Except for this time stretch of the experimental data the simulations reproduce the kinetics very well. Additionally, an existing Schottky barrier at the gold-\skut interface would amplify this selectivity between transport between 800 nm and 400 nm. To date we can not confirm the effect of a Schottky barrier. We have also investigated Yb-doped \skut films \cite{daniel16}, which show modified electrical conductivity \cite{daniel14}, or a change in band structure \cite{isaacs19}. These Yb-filled films essentially show the same behaviour of preferentially heating the \skut layer as well as fluence-dependent heat distribution at 400 nm excitation.

\section{Conclusions}

We have studied the thermal dynamics of a gold-skutterudite layer system upon photo-excitation of the top gold layer. The results show that while the majority of the laser energy is dissipated in the gold layer, the \skut layer get predominantly heated at short time delays after laser excitation. Excitation that includes interband transitions with 400 nm light shows this effect only after an increased laser fluence. 

The comparison to simulations according to the two-temperature model of interacting thermal electron and phonon subsystems suggests that the energy injection into the \skut layer would include non-thermal or ballistic electron migration from gold.  The inclusion of an electron-phonon coupling factor G, as reported in literature does not reproduce that strong relative heating of the buried \skut layer.  The occurrence of transport of non-thermal electrons across a nanoscale system would modify apparent electron relaxation rates, which could be interpreted as changed coupling factor in the TTM. Such ambiguity is not easily resolved in optical spectroscopy. In TDXTS, on the other hand, we follow a caloric approach, which returns the energy partitioning between the different subsystem, which enables to detect the ballistic transport, while not being sensitive to the electron subsystem directly.

The results show that ultrafast injection of charge carriers can be a dominant effect in conductive layered systems, in particular, if the laser-excited layer is characterized by a weak electron-phonon coupling. Contrary to the lower nominal optical penetration depth at 800 nm as compared to 400 nm the effect of heating the buried \skut layer at 800 nm laser wavelength is more pronounced, while the excitation at 400 nm shows a more distinct change with increasing external fluence. This is interpreted as a difference in electron temperature with different type of excitation of interband electrons (800 nm) or across the d to sp band for 400 nm. 

The role of different excitations is still not fully understood, but the study may open a tool to further clarify the dynamics. 

\section*{Appendix A: Thin film anisotropic thermal expansion}

Based on the Stoney equation for out-of-plane thermal stress due to differing thermal expansion coefficients of a continuous film on a substrate we formulate the linear coefficient of thermal expansion in the homogeneous case (no temperature gradients) as:

\begin{equation}
\frac{\Delta d}{d_0} = \frac{(\alpha_f - \alpha_s)}{1-\mu_f}\cdot \Delta T_{ph}
\end{equation}

that links the measured relative lattice expansion  $\Delta d/d_0$ with d$_0$ being the ambient lattice parameter to the difference in expansion coefficients between film $\alpha_f$ and substrate $\alpha_s$ and the Poisson ratio of the film $\mu_f$. The change in temperature is given by $\Delta T$. Similar relations have been described in \cite{suh88,zoo06,pudell18} and verified in \cite{plech19nanomat}. In this equation any plastic relaxation, in-plane expansion \cite{issenmann13} or substrate bending is excluded. As silicon as substrate shows a low CTE as compared to many films, such as gold or \skut (see table in Appendix B), the film will expand in excess of the bulk thermal expansion perpendicular to the surface, depending on the Poisson ratio.  The calculated, homogeneous CTE for gold will thus be 19.6$\cdot$10$^{-6}$/K, as compared to the measured value of 18.3 $\cdot$ 10$^{-6}$/K. CTE for \skut is calculated to be 11 $\cdot$ 10$^{-6}$/K,  as compared to the measured value of 8.3 $\cdot$ 10$^{-6}$/K . In the time-resolved case in this report the silicon substrate will not be heated significantly due to the heat dissipating faster than the inflow from the gold-skutterudite layer. Thus, the CTE will be higher by 30\% for gold and 50 \% for \skut, respectively.

\section*{Appendix B: Material parameters as input for NTMPy}

\begin{table}
\caption{\label{params} Parameters used in raw data analysis and solution of the TTM in NTMpy}
\footnotesize
\begin{tabular}{@{}llll}
\br
parameter & value & units &reference\\
\mr
gold film thickness d$_{Au}$     & 26           & nm                              & from XRR, fig. \ref{fig1}\\ 
gold refractive index 400 nm$^c$ & 1.59 + 1.91i &                                 & \cite{magnozzi19opt}\\       
gold refractive index 800 nm$^c$ & 0.13 + 5.04i &                                 & \cite{magnozzi19opt}\\       
gold density $\rho_{Au}$         & 19.3         & \gcm                            & \cite{lide95}\\   
gold CTE at 300 K                & 14.2$\cdot$10$^{-6}$ &/K                         & \cite{touloukian} \\
gold Poisson ratio               & 0.45         &                                 & \cite{goldmat} \\
gold el.$^a$ conductivity        & 315          & W/(m$\cdot$K)                   & \cite{huetter09}\\ 
gold ph.$^a$ conductivity        & 2.6          & W/(m$\cdot$K)                   & \cite{huetter09}\\ 
gold el. heat capacity $^d$      & 67/$\rho_{Au}\cdot \mbox{T}_e \/ ^b$
                                 & J/(m$^3\cdot \mbox{K}^2$)                      & \cite{lin08}\\ 
gold ph. heat capacity $^d$      & 2.7$\cdot$10$^6$/$\rho_{Au}$ & J/(m$^3\cdot$K) & \cite{huetter09}\\ 
gold e.-ph. coupling   $^d$      & 2$\cdot$10$^{16}$            & W/(m$^3\cdot$K) & \cite{lin08} \\ 
\skut film thickness d$_{skut}$  & 37           & nm                              & from XRR, fig. \ref{fig1}\\ 
\skut refractive index 400 nm    & 1.73+0.56i   &                                 & \cite{kiarii18}\\ 
\skut refractive index 800 nm    & 5.45+0.22i   &                                 & \cite{kiarii18}\\
\skut density                    & 7.62         & \gcm                            & \cite{caillat96}\\  
\skut CTE at 300 K$^{*}$         & 9.6$\cdot$10$^{-6}$ &/K                          & \cite{schupp03} \\
\skut Poisson ratio              & 0.22         &                                 & \cite{skutmat}\\
\skut el. conductivity$^{*}$     & 0.2          & W/(m$\cdot$K)                   & \cite{yang02skut}\\ 
\skut ph. conductivity$^{*}$     & 8            & W/(m$\cdot$K)                   & \cite{yang02skut}\\ 
\skut el. heat capacity          & 0.95/$\rho_{CoSb3}\cdot$T$_e$ $^b$
                                 & J/(m$^3\cdot$K$^2$)                            & \cite{caillat96}\\ 
\skut ph. heat capacity          & 2.6  $\cdot$10$^6$/$\rho_{CoSb3}$ 
                                 & J/(m$^3\cdot$K    )                            & \cite{zhang10skut}\\
\skut e.-ph. coupling            & 7$\cdot$10$^{16}$& W/(m$^3\cdot$K)               & \cite{yang02skut} \\
SiO$_2$ film thickness d$_{SiO2}$& 100          & nm                              & from growth\\
SiO$_2$ refractive index 400 nm  & 1.47         &                                 & \cite{arosa20}$^c$ \\
SiO$_2$ refractive index 800 nm  & 1.45         &                                 & \cite{arosa20}\\
SiO$_2$ density                  & 2.2          & \gcm                            & \cite{lide95} \\
SiO$_2$ el. conductivity         & 0            &                                 &   \\
SiO$_2$ ph. conductivity         & 10.5         & W/(m$\cdot$K)                   & \cite{cahill04}\\
SiO$_2$ el. heat capacity        & 0            & J/(m$^3\cdot$K)                 & \\
SiO$_2$ ph. heat capacity        & 2.2          & J/$\rho_{SiO2}$/(m$^6 \cdot$K)& \cite{lide95}\\
SiO$_2$ e.-ph. coupling          & 0            &                                 &  \\
Si refractive index 400 nm       & 5.56 +0.38i  &                                 & \cite{aspnes83}\\
Si refractive index 800 nm       & 3.69 + 0.007i&                                 & \cite{aspnes83}\\
Si density                       & 2.33         & \gcm                            & \cite{lide95} \\
Si CTE at 300 K                  & 3.4$\cdot$10$^{-6}$ &/K                          & \cite{okada84} \\
Si el. conductivity$^{*}$        & .1           & W/(m$\cdot$K)                   &  \cite{glassbrenner64} \\
Si ph. conductivity              & 140          & W/(m$\cdot$K)                   &  \cite{glassbrenner64} \\
Si el. heat capacity             & 150/$\rho_{Si}$*T$_e$         & J/(m$^3\cdot$K$^2$)& \\
Si ph. heat capacity             & 1.65$\cdot$10${^6}$/$\rho_{Si}$ & J/(m$^3\cdot$K$^2$) & \cite{lide95}\\
Si e.-ph. coupling               & 0            &                                 &  \\
incidence angle of laser         &  1.257        & rad                             &\\

\br
\end{tabular}\\
$^{a}$ e. - electron, ph. - phonon; 

$^{b}$ T$_e$ electron temperature; 

$^c$ see refractive index compilation at https://refractiveindex.info/ ;

$^d$ see compilation of properties of gold at https://compmat.org/Resources.html.

$^*$ varies with doping level \cite{caillat96,sofo98}.

\end{table}
\normalsize

\clearpage

\section*{Appendix C: laser absorption profiles in the gold layer}

The laser absorption profiles within the gold layer have been isolated from the NTMPy simulations using the values in Appendix B, but setting the  electron heat conductivities  and electron-phonon coupling in gold and \skut to zero. The pulse length has been chosen to 100 fs after which the temperature profile is extracted at low fluence (15 \Jms for 400 nm and 150 \Jms  for 800 nm).

\begin{figure}
\begin{center}
\epsfig{file=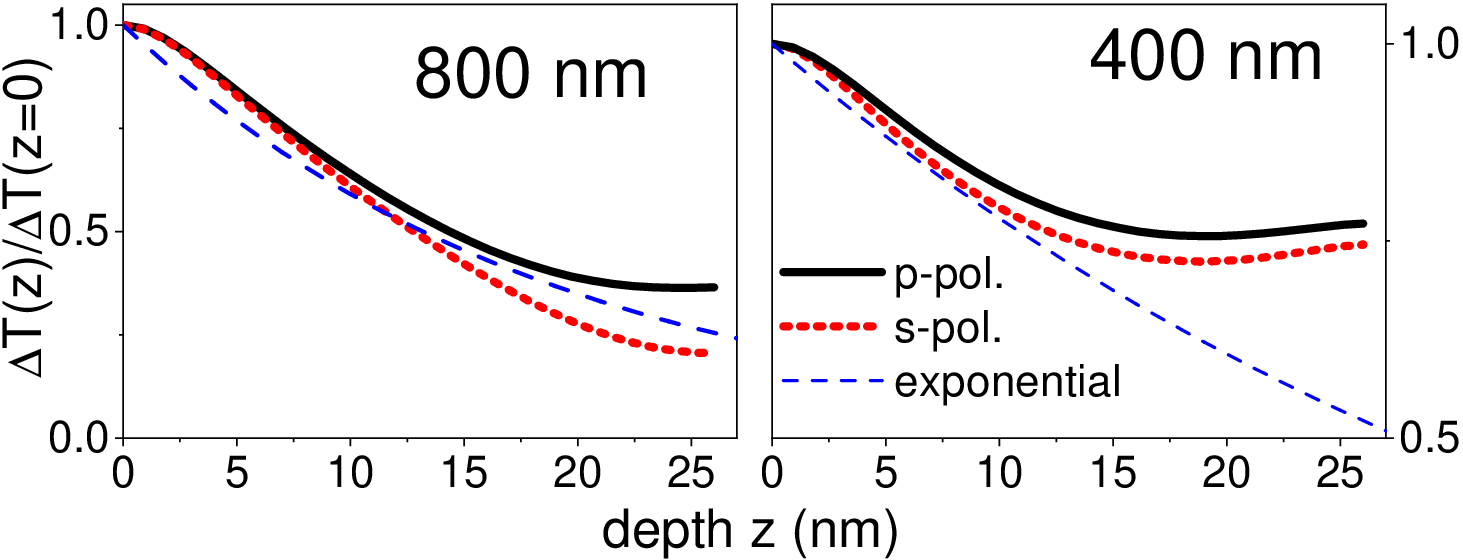,width=12cm, clip=} \caption{Results of the electron temperature depth profiles of the NTMPy simulation of the excitation of the layer system at 400 and 800 nm with 100 fs laser pulses and electron and phonon transport switched off. The dashed blue lines are guides to the eye of an exponential decay with 28.5 nm and 39 nm penetration depth, for 800 nm and 400 nm excitation, respectively.}
 \label{fig8}
 \end{center}
 \end{figure}

\section*{Data availability statement}

All data and Python-based codes can be made available upon reasonable request.

\section*{Acknowledgments}
This work is supported by the research in the program "Matter" in the topic "from Natter to Materials and Life (MML)" by the Helmholtz association. Beamtime at the ESRF and the KIT Light source is acknowledged. We wish to thank the institute IBPT at KIT for operation of the accelerator KARA at KIT. Initial part of the experiments were funded within the priority program 1364 "Nanostructured Thermoelectrics" of the German Research Foundation DFG.  
\section*{ORCID iDs}

\section*{References}
\providecommand{\newblock}{}


\begin{thebibliography}{10}
\expandafter\ifx\csname url\endcsname\relax
  \def\url#1{{\tt #1}}\fi
\expandafter\ifx\csname urlprefix\endcsname\relax\def\urlprefix{URL }\fi
\providecommand{\eprint}[2][]{\url{#2}}

\bibitem{shockley61}
Shockley W and Queisser H~J 1961 {\em J. Appl. Phys.\/} {\bf 32} 510

\bibitem{ross82}
Ross R~T and Nozik A~J 1982 {\em J. Appl. Phys.\/} {\bf 53} 3813–3818

\bibitem{tagliabue18}
Tagliabue G, Jermyn A~S, Sundararaman R, Welch A~J, DuChene J~S, Pala R,
  Davoyan A~R, Narang P and Atwater H~A 2018 {\em Nature Comm.\/} {\bf 9} 3394

\bibitem{su23}
Su Z~C, Chang C~H, Jhou J~C, Lin H~T and Lin C~F 2023 {\em Scientific
  Reports\/} {\bf 13} 5388

\bibitem{hohlfeld00}
Hohlfeld J, Wellershoff S~S, Gudde J, Conrad U, Jahnke V and Matthias E 2000
  {\em Chem. Phys.\/} {\bf 251} 237--258

\bibitem{pudell18}
Pudell J, Maznev A~A, Herzog M, Kronseder M, Back C~H, Malinowski G, von
  Reppert A and Bargheer M 2018 {\em Nature Comm.\/} {\bf 9} 3335

\bibitem{liu05}
Liu X, Stock R and Rudolph W 2005 {\em Phys. Rev. B\/} {\bf 72} 195431

\bibitem{lejman14}
Lejman M, Shalagatskyi V, Kovalenko O, Pezeril T, Temnov V~V and Ruello P 2014
  {\em J. Opt. Soc. Am. B\/} {\bf 31} 282--290

\bibitem{du13}
Du L, Furube A, Hara K, Katoh R and Tachiya M 2013 {\em J. Photochem.
  Photobiol. C\/} {\bf 15} 21–30

\bibitem{minutella17}
Minutella E, Schulz F,  and Lange H 2017 {\em J. Phys. Chem. Lett.\/} {\bf 8}
  4925--4929

\bibitem{karna23}
Karna P, Hoque M~S~B, Thakur S, Hopkins P~E and Giri A 2023 {\em Nano Lett.\/}
  {\bf 23} 491--496

\bibitem{choi14}
Choi G~M, Wilson R~B and Cahill D~G 2014 {\em Phys. Rev. B\/} {\bf 89} 064307

\bibitem{jang20}
Jang H, Kimling J and Cahill D~G 2020 {\em Phys. Rev. B\/} {\bf 101} 064304

\bibitem{wang12}
Wang W and Cahill D~G 2012 {\em Phys. Rev. Lett.\/} {\bf 109} 175503

\bibitem{li19}
Li W, Wang J, Xie Y, Gray J~L, Heremans J~J, Kang H~B, Poudel B, Huxtable S~T
  and Priya S 2019 {\em Chem. Mater.\/} {\bf 31} 862--872

\bibitem{pang24}
Pang X, He M, Zhang F, Jia B, Wang W, Cao X, Song M, Chao X, Yang Z and Wu D
  2024 {\em Chem. Eng. J.\/} {\bf 481} 148457

\bibitem{daniel16}
Daniel M~V, Lindorf M and Albrecht M 2016 {\em J. Appl. Phys.\/} {\bf 120}
  125306

\bibitem{he14}
He C, Daniel M, Grossmann M, Ristow O, Brick D, Schubert M, Albrecht M and
  Dekorsy T 2014 {\em Phys. Rev. B\/} {\bf 89} 174303

\bibitem{bracht14}
Bracht H, Eon S, Frieling R, Plech A, Issenmann D, Wolf D, {Lundsgaard Hansen}
  J, {Nylandsted Larsen} A, {Ager III} J and Haller E~E 2014 {\em New J.
  Phys.\/} {\bf 16} 015021

\bibitem{plech19nanomat}
Plech A, Krause B, Baumbach T, Zakharova M, Eon S and Bracht H 2019 {\em
  nanomaterials\/} {\bf 9} 501

\bibitem{daniel14}
Daniel M 2014 {\em Structural and Thermoelectric Properties of Binary and
  Ternary Skutterudite Thin Films\/} Ph.D. thesis TU Chemnitz

\bibitem{genx}
Bjorck M and Andersson G 2007 {\em J. Appl. Cryst.\/} {\bf 40} 1174

\bibitem{touloukian}
Touloukian Y~S, Kirby R~K, Taylor R~E and Desai P~D 1975 {\em Thermal expansion
  - Metallic elements and alloys\/} vol~12 (IFI Plenum, New York)
  thermodynamical properties of matter ed

\bibitem{plech02jsr}
Plech A, Randler R, Geis A and Wulff M 2002 {\em J. Synchrotron Rad.\/} {\bf 9}
  287--292

\bibitem{cammarata09}
Cammarata M, Eybert L, Ewald F, Reichenbach W, Wulff M, Anfinrud P, Schotte F,
  Plech A, Kong Q, Lorenc M, Lindenau B, R\"abiger J and Polachowski S 2009
  {\em Rev. Sci. Instr.\/} {\bf 80} 15101

\bibitem{cahill03}
Cahill D~G, Ford W~K, Goodson K~E, Mahan G~D, Majumdar A, Maris H~J, Merlin R
  and Phillpot S~R 2003 {\em J. Appl. Phys.\/} {\bf 93} 793

\bibitem{chen99}
Chen G and Hui P 1999 {\em Thin Solid Films\/} {\bf 339} 58

\bibitem{issenmann13}
Issenmann D, Eon S, Wehmeier N, Bracht H, Buth G, Ibrahimkutty S and Plech A
  2013 {\em Thin Solid Films\/} {\bf 541} 28

\bibitem{alber21}
Alber L, Scalera V, Unikandanunni V, Schick D and Bonetti S 2021 {\em Computer
  Phys. Comm.\/} {\bf 265} 107990

\bibitem{bonn00}
Bonn M, Denzler D~N, Funk S and Wolf M 2000 {\em Phys. Rev. B\/} {\bf 61}
  1101--1106

\bibitem{chen11pnas}
Chen J, Chen W~K, Tang J and Rentzepis P~M 2011 {\em Proc. Natl. Acad. Sci.
  USA\/} {\bf 108} 18887--18892

\bibitem{fot19}
Foteinopoulou S, Devarapu G~C~R, Subramania G~S, Krishna S and Wasserman D 2019
  {\em Nanophotonics\/} {\bf 8} 2129--2175

\bibitem{perner97}
Perner M, Bost P, Lemmer U, von Plessen G, Feldmann J, Becker U, Mennig M,
  Schmitt M and Schmidt H 1997 {\em Phys. Rev. Lett.\/} {\bf 78} 2192

\bibitem{bracht12}
Bracht H, Wehmeier N, Eon S, Plech A, Issenmann D, {Lundsgaard Hansen} J,
  {Nylandsted Larsen} A, {Ager III} J and Haller E 2012 {\em Appl. Phys.
  Lett.\/} {\bf 101} 064103

\bibitem{homola99}
Homola J, Yee S and Gauglitz G 1999 {\em Sens. Actuators B Chem.\/} {\bf 54}
  3--15

\bibitem{chiang23}
Chiang W~Y, Bruncz A, Ostovar B, Searles E~K, Brasel S, Hartland G and Link S
  2023 {\em J. Phys. Chem. C\/} {\bf 127} 21176--21185

\bibitem{hopkins10}
Hopkins P~E 2010 {\em J. Heat Transfer\/} {\bf 132} 014504

\bibitem{isaacs19}
Isaacs E~B and Wolverton C 2019 {\em Chem. Mater.\/} {\bf 31} 6154--6162

\bibitem{suh88}
Suh I~K, Ohta H and Waseda Y 1988 {\em J. Mat. Sci.\/} {\bf 23} 757--760

\bibitem{zoo06}
Zoo Y, Adams D, Mayer J and Alford T 2006 {\em Thin Solid Films\/} {\bf 513}
  170--174

\bibitem{magnozzi19opt}
Magnozzi M, Ferrera M, Mattera L, Canepa M and Bisio F 2019 {\em Nanoscale\/}
  {\bf 11} 1140--1146

\bibitem{lide95}
Lide D~R 2015 {\em CRC Handbook of Chemistry and Physics\/} 95th ed (CRC Press,
  Boca Raton, FL)

\bibitem{goldmat}
Persson K 2016 Materials data on {Au (SG:225)} by {Materials Project}

\bibitem{huetter09}
Huettner B 2009 {\em Femtosecond Laser Pulse Interactions with Metals\/} vol~xx
  (Springer) pp 315--337 the theory of laser materials processing ed

\bibitem{lin08}
Lin Z, Zhigilei L~V and Celli V 2008 {\em Phys. Rev. B\/} {\bf 77} 075133

\bibitem{kiarii18}
Kiarii E~M, Govender K~K, Mamo M~A and Govender P~P 2018 {\em
  ChemistrySelect\/} {\bf 3} 9336--9347

\bibitem{caillat96}
Caillat T, Borshchevsky A and Fleurial J~P 1996 {\em J. Appl. Phys.\/} {\bf 80}
  4442--4449

\bibitem{schupp03}
Sch\"upp B, B\"acher I, Hecker M, Mattern N, Savchuk V and Schumann J 2003 {\em
  Thin solid films\/} {\bf 434} 75--81

\bibitem{skutmat}
Persson K 2014 Materials data on {CoSb$_3$} ({SG}:204) by materials project

\bibitem{yang02skut}
Yang J, Morelli D~T, Meisner G~P, Chen W, Dyck J~S and Uher C 2002 {\em Phys.
  Rev. B\/} {\bf 65} 094115

\bibitem{zhang10skut}
Zhang Y, Li C, Xu G, Du Z, Guo C and Li J 2010 {\em Int. J. Mat. Res. (formerly
  Z. Metallkd.)\/} {\bf 101} 808--811

\bibitem{arosa20}
Arosa Y and {de la Fuente} R 2020 {\em Opt. Lett.\/} {\bf 45} 4268--4271

\bibitem{cahill04}
Cahill D~G 2004 {\em Rev. Sci. Instr.\/} {\bf 75} 5119--5123

\bibitem{aspnes83}
Aspnes D~E and Studna A~A 1983 {\em Phys. Rev. B\/} {\bf 27} 985--1009

\bibitem{okada84}
Okada Y and Tokumaru Y 1984 {\em J. Appl. Phys.\/} {\bf 56} 314--320

\bibitem{glassbrenner64}
Glassbrenner C~J and Slack G~A 1964 {\em Phys. Rev.\/} {\bf 134} A1058

\bibitem{sofo98}
Sofo J~O and Mahan G~D 1998 {\em Phys. Rev. B\/} {\bf 58} 15620--15623

\end{thebibliography}
\end{document}